\documentstyle[11pt,epsfig]{article}
\begin{document}

\title{Strangeness Electromagnetic Production on Nucleons and Nuclei}
\author{Petr Byd\v{z}ovsk\'y\footnote{bydz@ujf.cas.cz} and Miloslav Sotona\\
{\it Nuclear Physics Institute, \v{R}e\v{z} near Prague, Czech Republic}}
\maketitle

\begin{abstract}
 Isobar models for the electromagnetic production of kaons are discussed 
with emphasis on the K$^+$ photoproduction at very small kaon angles and 
K$^0$ photoproduction on deuteron. Distorted-wave impuls approximation 
calculations of the cross sections for the electroproduction of hypernuclei 
are presented on the case of the $^{12}_{\ \Lambda}$B production.  
\end{abstract}

{\it Key words:} Kaon, photoproduction, electroproduction, hypernucleus

{\it PACS:} 13.60.Le, 25.20.Lj, 25.30.Rw, 21.80.+n

\section{Introduction}
\label{intr}
The photo- and electroproduction of kaons on nucleons and nuclei provides 
us with additional information about a structure and interactions of baryons.
Studying the production on nucleons we learn about a reaction mechanism 
(quark models versus an effective Lagrangian theory with hadrons), couplings 
and form factors of hadrons, and look for new ``missing'' resonances. 
Moreover, we also need a correct description of the elementary production 
process, particularly at small kaon angles, to study the electroproduction 
of hypernuclei~\cite{ProdH}.

Producing a $\Lambda$ hyperon inside ordinary nuclei allows us to probe 
a structure deep inside the nucleus because the $\Lambda$s are not Pauli 
blocked there. Revealing the structure of hypernuclei gives us new information 
about the hyperon-nucleon interactions, especially about the spin-dependent 
part which is not accessible by today viable experiments on the 
$\Lambda$-nucleon scattering. 

A detailed study of the electromagnetic strangeness production is enabled 
by new ample data of good quality from various laboratories: JLab, ELSA, 
SPring-8, ESRF, LNS and MAMI. These data allow to perform a detailed 
analysis of the elementary process~\cite{MartSul,ByMa}. Recent data on 
the photoproduction of neutral kaons off deuteron from Tohoku 
University~\cite{Wat07,Tsu08} facilitate doing more rigorous tests 
of models for the elementary process~\cite{Tsu08,Salam09}.
 
\section{Photo- and Electroproduction of Kaons on Nucleons}
\label{elem}

Description of the electroproduction process can be formally reduced 
to investigation of a binary process of the photoproduction by virtual 
photons since the electromagnetic coupling constant is small enough 
to justify the one-photon exchange approximation. 

There are many approaches to treat the photoproduction process. 
Among them isobar models based on an effective Lagrangian description 
considering only hadron degrees of freedom are suitable for their 
further use in the more complex calculations of the electroproduction 
of hypernuclei. Other approaches are eligible either for higher energies 
($E_{\gamma}>4$ GeV), the Regge model~\cite{Regge}, or to the 
threshold region, the Chiral Perturbation Theory~\cite{ChPT}. 
Quark models~\cite{SagQM} are too complicated for their further 
use in the hypernuclear calculations. Another approach, aimed 
at the forward-angle production, is the hybrid Regge-plus-resonance 
model~\cite{RPR} (RPR) in which a nonresonant part of the amplitude is 
generated by the $t$-channel Regge-trajectory exchange and a resonant 
behaviour is shaped by the exchanges of s-channel resonances like 
in isobar models. 

In the effective hadron Lagrangian approach, various channels 
connected via the final-state interaction (meson-baryon 
rescattering processes) have to be treated simultaneously to take 
unitarity properly into account~\cite{CouCh,Chi01}.  
In Ref.~\cite{Chi01} the coupled-channel approach has been 
used to include effects of the $\pi$N intermediate states in 
the p($\gamma$,K$^+$)$\Lambda$ process. However, tremendous 
simplifications originate in neglecting the rescattering effects 
in the formalism assuming that they are included to some extent 
by means of effective values of the strong coupling constants 
fitted to data. This simplifying assumption was adopted in many 
of the isobar models, e.g., Saclay-Lyon A (SLA)~\cite{SLA}, 
Kaon-MAID (KM)~\cite{KM}, and Janssen {\it et al.}~\cite{Jan01}.

In the isobar models, the amplitude obtains contributions from 
the Born terms and exchanges of resonances. Due to absence of 
a dominant resonance in the photoproduction of kaons (unlike 
in the $\pi$ and $\eta$ photoproduction) large number of various  
combinations of the resonances with mass below 2 GeV must be 
taken into account~\cite{SLA}. This number of models (resonance 
combinations) is limited considering constraints set by 
SU(3)~\cite{SLA,KM} and crossing symmetries~\cite{WJC92,SLA} and 
by duality hypothesis~\cite{WJC92}. 
Adopting the SU(3) constraints to the two main coupling constants, 
however, makes the contribution of the Born terms nonphysically 
large~\cite{Jan01}. To reduce this contribution, either hyperon 
resonances~\cite{SLA} or hadron form factors~\cite{KM} must 
be added, or a combination of both~\cite{Jan01}. Here we use 
especially the SLA and KM models. 

Common to the KM and SLA models is that, besides the extended Born 
diagrams, they also include the kaon resonances $K^*(890)$ and $K_1(1270)$.  
These models differ in a choice of $s$- and $u$-channel resonances 
in the intermediate state, in a treatment of the hadron 
structure, and in a set of experimental data to which free 
parameters were adjusted. However, the two main coupling constants,
$g_{KN\Lambda}$ and $g_{KN\Sigma}$, fulfil  limits of 20\% broken
SU(3) symmetry \cite{SLA} in both models. 
In the SLA model, one  nucleon and four hyperon resonances are included 
and their coupling constants were fitted to the old data~\cite{AS1} and 
the first results of SAPHIR by Bockhorst {\it et al.} \cite{SAPHIR94}. 
In the KM model, four nucleon but no hyperon resonances were assumed and 
parameters of the model were fitted to the old and SAPHIR \cite{SAPHIR98} 
data. The SLA and KM models were expected to provide reasonable results 
for the photon energies below 2.2 GeV. 
In the SLA model, hadrons are treated as point-like objects, in contrast 
to the KM model in which phenomenological hadron form factors are inserted 
in the hadron vertexes \cite{KM} to stimulate a structure of hadrons.
The form factors are included maintaining the gauge invariance of the 
amplitude~\cite{KM}.
     
\subsection{Forward-angle K$^+\Lambda$ Photoproduction}
\begin{figure}[ht!]
\centerline{\psfig{figure=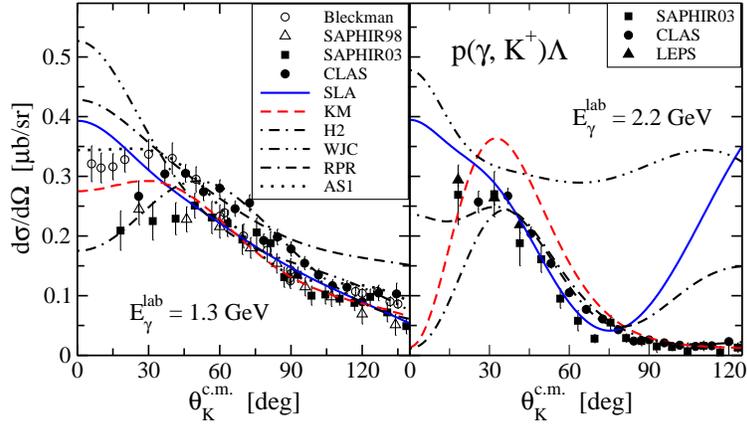,angle=270,width=10.cm}}
\caption{Predictions of the models Saclay-Lyon A~\cite{SLA}, 
Kaon-MAID~\cite{KM}, H2~\cite{HYP03}, Williams-Ji-Cotanch (WJC)~\cite{WJC92}, 
Regge-Plus-Resonance (RPR)~\cite{RPR}, and Adelseck-Saghai (AS1)~\cite{AS1} 
are compared with data from Refs.~\cite{Bleck}~(Bleckman), 
\cite{SAPHIR98}~(SAPHIR98), \cite{SAPHIR03}~(SAPHIR03), \cite{CLAS05}~(CLAS), 
and \cite{LEPS}~(LEPS) at two photon laboratory energies.} 
\label{fig1}
\end{figure}

A good quality description of the elementary process minimises uncertainty 
in the calculations of excitation spectra for the hypernucleus 
electroproduction~\cite{ProdH}. In Fig.~\ref{fig1} we compare predictions 
of various phenomenological models for the differential cross section with 
experimental data at two photon laboratory energies. 
For the lower energy (the left panel) the theoretical predictions reveal 
a good agreement with data for $\theta_{\rm K}^{\rm c.m.} > 45^{\circ}$. 
Results for the kaon angles smaller than $45^{\circ}$ differ considerably 
even for the modern models SLA and KM. These models differ by~$\approx 30$\% 
at $\theta_{\rm K}^{\rm c.m.} \approx 0$. At higher energy, 2.2 GeV 
(the right panel), the discrepancy of the model predictions at forward kaon 
angles is as large as several hundreds of per-cent.
%
% Figure 1
%  

The damping of the cross sections at very forward angles and higher 
energies, seen in the right panel of Fig.~\ref{fig1}, can be understood 
based on analysis of the SLA and H2 \cite{HYP03} models. 
In the SLA model, the cross section at $\theta_{\rm K}^{\rm c.m.} < 30^{\circ}$ 
is dominated by the Born terms, particularly by the electric part of 
the proton exchange. This contribution makes a plateau in the cross 
section as the angle goes to zero. This feature would be similar in the H2 
model since values of the coupling constants, $g_{\rm K^+\Lambda p}$ and 
$g_{\rm K^+\Sigma p}$,  are almost equal to those in SLA. The situation is, 
however, different due to a strong suppression of the proton term caused 
by the hadron form factors in the H2 model, which is of the order 
of 0.002 in the cross section at 2 GeV. In this energy region, 
resonances contribute only in interference with the Born terms so 
that they cannot fill up the dip. This strong suppression of the Born 
terms in the models with the hadron form factors leads, therefore, to 
a pronounced dip in the differential cross section at very small angles 
(see Fig.~\ref{fig1} for the results of the H2 and KM models). Lack and 
mutual inconsistency of data at very small angles 
makes fitting of models in this kinematical region problematic~\cite{ByMa}.

%
% Figure 2
%  
\begin{figure}[b!]
\centerline{\psfig{figure=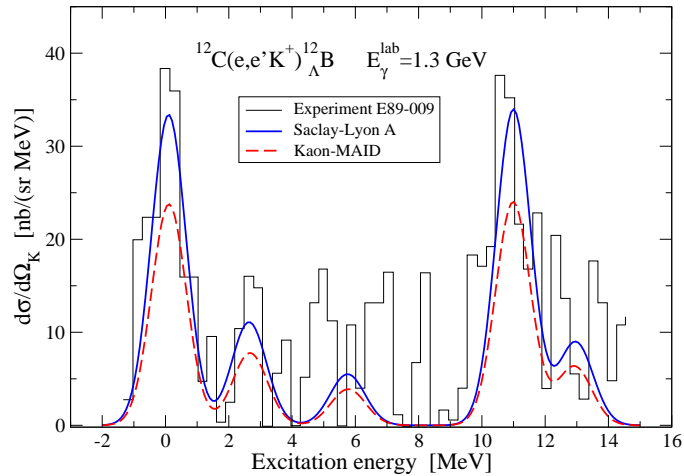,angle=270,width=9.cm}}
\caption{Results of the DWIA calculations of the cross section for the 
electroproduction of $^{12}_{\ \Lambda}$B at 1.3 GeV using the Saclay-Lyon A 
and Kaon-MAID models for the elementary production operator. 
Data are adopted from Ref.~\cite{E89-009}.} 
\label{fig2}
\end{figure}
This low-quality description of the elementary process in the kinematical 
region relevant for the hypernucleus calculations causes a large uncertainty 
in predictions of the cross sections for the hypernucleus production 
as demonstrated in Fig.~\ref{fig2} for the energy 1.3 GeV. 
Theoretical calculations of the cross sections for the electroproduction 
of $^{12}_{\ \Lambda}$B (the small momentum transfer,  
$Q^2 = -3\cdot 10^{-6}$ (GeV/c)$^2$, allows a comparison with 
the photoproduction cross section) are compared with the data from JLab 
experiment in Hall C~\cite{E89-009}. The difference in model predictions 
amounts to $\approx 30$\% which corresponds to the difference at  
$\theta_{\rm K}^{\rm c.m.} \approx 0$ seen in the left panel of Fig.~\ref{fig1}.  

\section{Photoproduction of K$^0$ on Deuteron}
\label{deut}
The inclusive cross section for the d($\gamma$,K$^0$)YN' or 
d($\gamma$,$\Lambda$)KN' reactions (Y stands for $\Lambda$ or $\Sigma$ 
and N' for proton or neutron) in a threshold region is calculated 
in the impulse approximation in which the nucleon N' acts as 
a spectator~\cite{K0d04}. Contributions of the final-state 
interaction (FSI) to the inclusive cross section were shown to be 
small in the studied kinematical region~\cite{Agus,Salam09}.  
Moreover, we assume that a part of the KY rescattering effects 
is absorbed in the coupling constants of the elementary amplitude 
and that the KN interaction is weak on the hadron scale. 
The main lack of precision, therefore, comes from ignoring 
the YN FSI. 
In the calculations we have chosen the {\it off-shell} 
approximation~\cite{K0d04} in which the four-momentum is conserved in 
the production vertex forcing the target nucleon off its mass shell. 
The momentum distribution of the target nucleon is described by the 
nonrelativistic Bonn deuteron wave function OBEPQ~\cite{Bonn}.
More details about the calculations can be found in Ref.~\cite{K0d04}.

In the elementary amplitude, the strong coupling constants in the 
K$^0\Lambda$ and K$^+\Lambda$ channels are related via the SU(2) isospin 
symmetry, e.g. $g_{{\rm K}^+\Lambda {\rm p}}= g_{{\rm K}^0\Lambda {\rm n}}$ 
and $g_{{\rm K}^+\Sigma^0 {\rm p}}= -g_{{\rm K}^0\Sigma^0 {\rm n}}$. 
In the electromagnetic vertexes a ratio of the neutral to charged 
coupling constants have to be known. For the nucleon and its resonances 
the ratio can be related to the known helicity amplitudes of the 
nucleons~\cite{PDP} whereas for the kaon resonances the ratio relates 
to the decay widths known only for the K$^*$ meson~\cite{PDP}: 
r$_{\rm K^*}=-\sqrt{\Gamma_{{\rm K}^{*0}\rightarrow {\rm K}^0\gamma}/ 
\Gamma_{{\rm K}^{*+}\rightarrow {\rm K}^+\gamma}}=-1.53$ where the sign 
was set from the quark model prediction. Since the decay widths of 
the K$_1$ meson are unknown the appropriate ratio,  r$_{\rm K1}$, 
have to be fixed in the models. It was fitted to the K$^0\Sigma^+$ 
data in KM~\cite{KM}, r$_{\rm K1}=-0.45$, but it is a free 
parameter in SLA (see also Refs.~\cite{Tsu08} and \cite{K0d04}). 
For hyperons, the electromagnetic vertex is not changed in the 
K$^0\Lambda$ channel.

%
% Figure 3
%  
\begin{figure}[b!]
\centerline{\psfig{figure=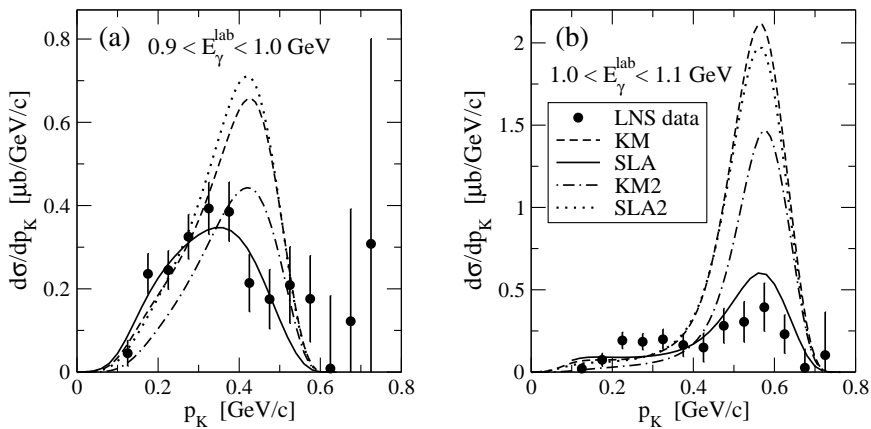,angle=270,width=11.5cm}}
\caption{Inclusive energy-averaged, (a) 0.9 $<$ E $<$ 1.0 GeV and 
(b)  1.0 $<$ E $<$ 1.1 GeV, and kaon-angle-integrated, 
0.9 $<$ cos($\theta_K$) $<$ 1.0, momentum spectra for the 
d($\gamma$,K$^0$)YN' reaction. Calculations with the $\Lambda$p in 
the final state using various elementary amplitudes are compared 
with data from Ref.~\cite{Tsu08}.} 
\label{fig3}
\end{figure}
In Figure~\ref{fig3} results for the momentum spectra of the 
d($\gamma$,K$^0$)$\Lambda$p reaction calculated with different elementary 
amplitudes are shown in comparison with recent data on the 
d($\gamma$,K$^0$)YN' reaction~\cite{Tsu08}. Contributions of the $\Sigma$ 
channels are very small in the low-energy region (Fig.~\ref{fig3}a) 
and at the spectator-kinematics region, the main peak in 
Fig.~\ref{fig3}b~\cite{Tsu08}. The KM model do not describe the data 
well, especially at the spectator-kinematics region (dashed line). 
Results of KM cannot be improved even if the r$_{\rm K1}$ parameter is 
refitted to the low-energy data (Fig.~\ref{fig3}a): $\chi^2_{\rm n.d.f.}= 3.64$ 
with r$_{\rm K1}= -2.34$. Results of this fit are displayed as KM2 
(dashed-dotted line) in Fig.~\ref{fig3}. It is seen that the KM model 
cannot fit a proper shape of the momentum spectra. 

The SLA model with r$_{\rm K1}= -2.09$, fitted to the low-energy data 
($\chi^2_{\rm n.d.f.}$= 0.88)~\cite{Tsu08}, gives good results in both 
energy regions. To demonstrate a flexibility of this model we show 
results of the model with r$_{\rm K1}= -3.4$ denoted SLA2. This version 
of the Saclay-Lyon model gives very similar values of the cross sections 
for the n($\gamma$,K$^0$)$\Lambda$ process as the KM model and therefore 
its results for the d($\gamma$,K$^0$)$\Lambda$p reaction are close to 
those of the KM model (Fig.~\ref{fig3}). This demonstrates that results of the 
SLA model in the K$^0$p channel are very sensitive to the ratio of the 
decay widths of the K$_1$ resonance ($r_{\rm K1}$).

The angular dependence of the elementary cross sections differs for the 
models KM and SLA. The latter gives the cross sections enhanced at the 
backward hemisphere whereas the KM model reveal a more complex resonance 
pattern dominated by the N$^*$(1720) resonance~\cite{Tsu08,K0d04}.
In general, the elementary models which possess the backward-peaked cross 
sections, e.g. SLA, give much better results for the 
d($\gamma$,K$^0$)$\Lambda$p inclusive cross sections than those with 
other angular dependence~\cite{Tsu08}, e.g. KM and SLA2 in 
Fig.~\ref{fig3}. This comparison shows that the Tohoku data prefer models 
which predict the cross section for the n($\gamma$,K$^0$)$\Lambda$ process  
enhanced at the backward hemisphere. 

\begin{figure}[hb]
\centerline{\psfig{figure=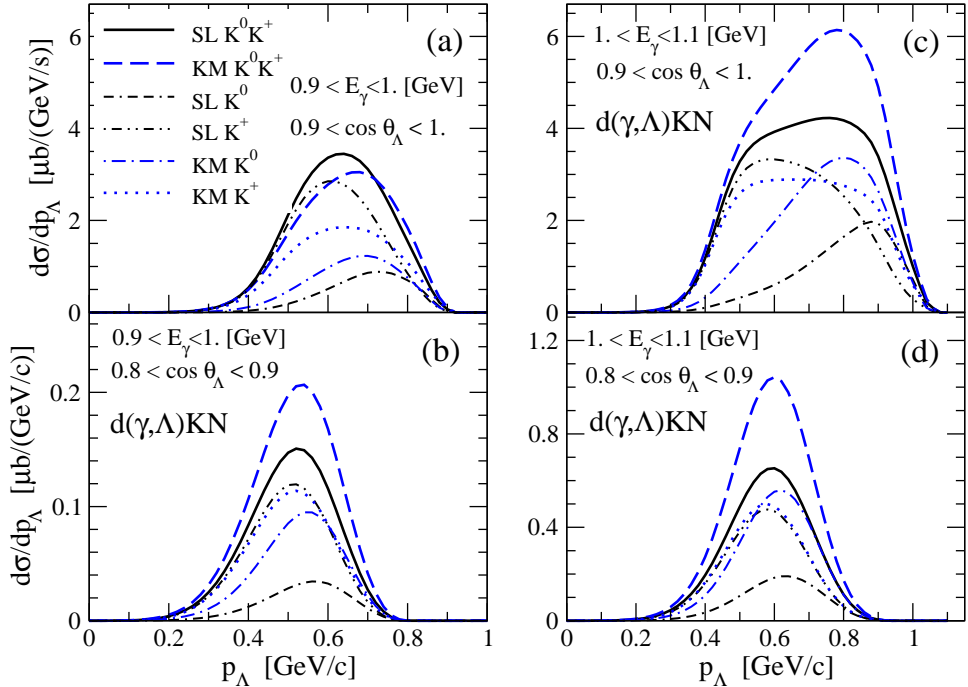,angle=270,width=13.cm}}\vspace{2mm}
\caption{Predictions of inclusive energy-averaged and 
$\Lambda$-angle-integrated momentum spectra for the 
d($\gamma$,$\Lambda$)KN' reaction. Results of the SLA and KM models 
are shown for various final states.} 
\label{fig4}
\end{figure}

Predictions of the momentum distributions in the d($\gamma$,$\Lambda$)KN'  
reaction are shown in Fig.~\ref{fig4}. Results for the energy averaged 
and $\Lambda$-angle integrated inclusive cross sections are plotted for 
the K$^0$p and K$^+$n channels using the SLA and KM models. Predictions 
of the models agree each other for the K$^+$n channel except for the 
low-energy and small-angle region (Fig.~\ref{fig4}a). In the kinematical 
regions of Figs.~\ref{fig4}b, c, and d, the full results (K$^+$n + K$^0$p) 
of the models differ enough to allow a new data to discriminate between 
the models.\vspace{2mm}  
%
% Figure 4
%  

%
% Figure 5
%  
\begin{figure}[ht]
\centerline{\psfig{figure=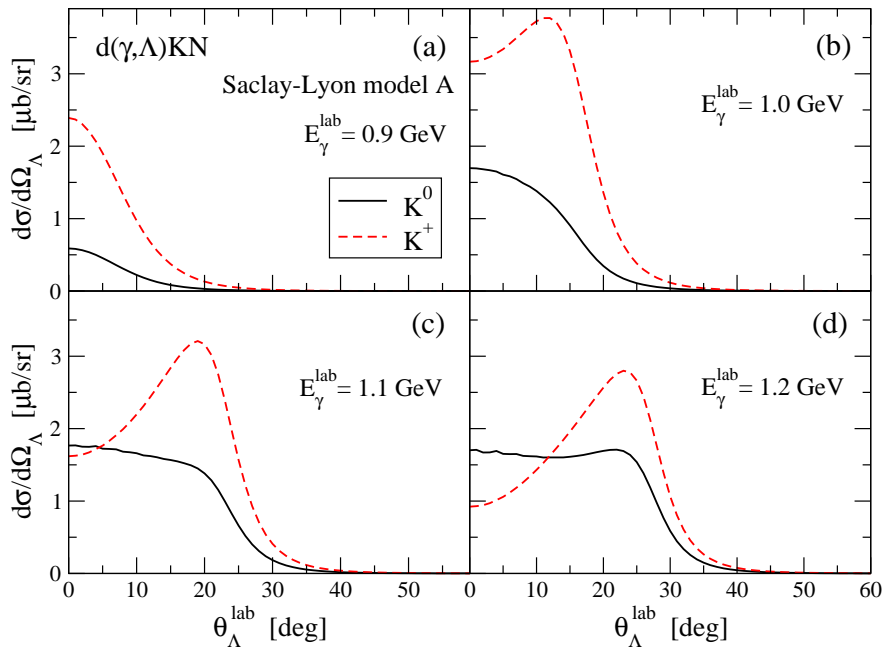,angle=270,width=12.cm}}
\caption{Predictions of the $\Lambda$-momentum-integrated cross sections as 
a function of $\Lambda$ laboratory angle. Calculations in PWIA using the 
Saclay-Lyon A model with r$_{\rm K1}=-2.09$ are shown for K$^0$ and K$^+$ 
in the final state.} 
\label{fig5}
\end{figure}
In Figure~\ref{fig5} the inclusive differential cross section is plotted  
in dependence on the $\Lambda$ angle. Results show that the cross 
sections for the K$^+$n and K$^0$p channels are centred at $\Lambda$ angles  
smaller than $\approx 30^{\circ}$ for photon energies below 1.1 GeV. 
This indicates a possibility to measure the total cross section of the 
reaction if an acceptance of the detector can cover this small-angle region. 
The reaction with the K$^+$n final state reveals different angular dependence 
than that with the K$^0$p state, being peaked at 20$^\circ$ for energy 
larger than 1 GeV (see Figs.~\ref{fig5}b, c, and d).

\section{Electroproduction of p-shell Hypernuclei}
\label{hyper}
The cross sections for the electroproduction of hypernuclei are 
calculated in the framework of the distorted-wave impulse approximation 
(DWIA)~\cite{SotFru}. 
The ingredients of the many-body matrix element in the DWIA are 
nonrelativistic nuclear and hypernuclear wave functions, an operator 
for the elementary-production process, an incoming virtual-photon wave 
function (in the one-photon exchange approximation), and an outgoing kaon 
distorted wave function. The calculations are performed in the laboratory 
frame neglecting the Fermi motion. For the elementary operator we use 
an isobar model, e.g. the Saclay-Lyon model discussed in Sect.~\ref{elem}. 
The kaon distorted wave is calculated in the optical-potential model 
assuming the eikonal approximation~\cite{SotFru}. 

The nuclear wave functions are obtained in the shell-model calculations 
with an effective p-shell interaction. The in-medium $\Lambda$N interaction 
used in the shell-model calculations of the hypernucleus wave functions 
is fitted to precise $\gamma$-ray spectra of p-shell nuclei~\cite{JohnM}.

\begin{figure}[ht!]
\centerline{\psfig{figure=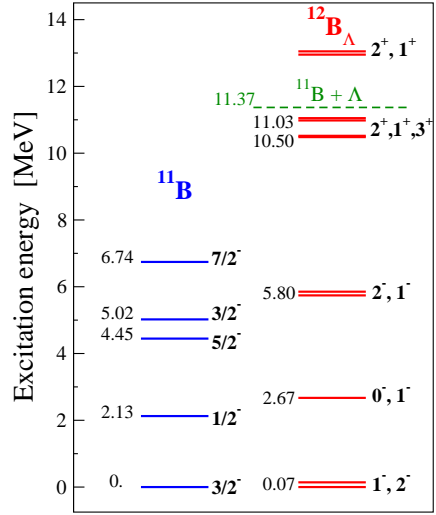,angle=270,width=6.1cm}}
\caption{Schemes of the low-lying energy levels for the core nucleus 
$^{11}$B (left) and hypernucleus $^{12}_{\ \Lambda}$B (right). The dashed 
line indicates the threshold for the $\Lambda$ emission. 
The energies are in MeV.\vspace{-2mm}} 
\label{fig6}
\end{figure}

In the right part of Fig.~\ref{fig6} we show low-energy excitation levels 
of the $^{12}_{\ \Lambda}$B build up on levels of the core nucleus $^{11}$B 
(the left part) assuming the weak coupling scheme. The low-energy states 
in the spectrum of $^{12}_{\ \Lambda}$B (doublets) are generated by coupling 
of the $\Lambda$ in $s$-state to the lowest negative-parity states in 
$^{11}$B. 
The higher-energy multiplets in $^{12}_{\ \Lambda}$B are formed by coupling 
of the $p_{1/2}$ or $p_{3/2}$ $\Lambda$'s to the lowest negative-parity 
states of $^{11}$B. The dashed line in Fig.~\ref{fig6} indicates the energy 
of the quasi-free production reaction. 
%
% Figure 6
%  

%
% Figure 7
%  
\begin{figure}[hb!]
\centerline{\psfig{figure=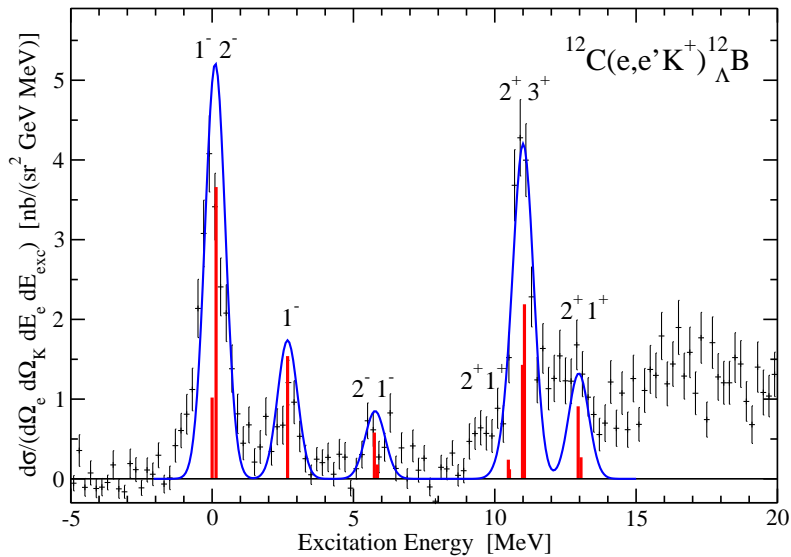,angle=270,width=10.5cm}}
\caption{Excitation-energy spectrum of $^{12}_{\ \Lambda}$B at 
E$_{\gamma}^{\rm lab}=2.2$ GeV calculated in DWIA using the SLA model. 
The overall prediction (solid line) and the cross sections for 
particular states (bars) are compared with data from Ref.~\cite{Mauro}.} 
\label{fig7}
\end{figure}
Results of the DWIA calculations using the SLA model for the 
electroproduction of $^{12}_{\ \Lambda}$B are compared in Fig.~\ref{fig7} 
with JLab data from Hall A~\cite{Mauro}. The cross sections 
for members of the multiplets are indicated by bars. Agreement of the 
calculations with data is good for the part of the spectra where 
the $\Lambda$ is in $s$-state (the fist three peaks).   
The region of $p_{\Lambda}$ states (the 2nd main peak), especially 
the strengths on both sides of the 2nd main peak, is not well described 
using these simple shell-model wave functions. More complex calculations 
have to be performed, e.g. assuming $1h\omega$ excitations, to better 
understand the $p_{\Lambda}$ region. Similar conclusions were also drawn 
for the $^{16}$O(e,e'K$^+$)$^{16}_{\ \Lambda}$N reaction~\cite{Cuss}. 
The reasonable values of the cross sections provided by the DWIA calculations 
with the SLA model suggest that the Saclay-Lyon model gives also good 
values of the elementary cross section at very forward kaon angles. 
 
\section{Summary}
Data at very small kaon angles are needed to fix the models for the K$^+$ 
photoproduction at forward angles which is necessary for reliable 
hypernuclear calculations. The data on the K$^0$ photoproduction off 
deuteron near threshold prefer the models which give enhancement of 
the cross section at the backward kaon angles.
Predictions of the DWIA shell-model calculations agree well with the 
spectra of $^{12}_{\ \Lambda}$B and $^{16}_{\ \Lambda}$N for the $\Lambda$ 
in $s$-state. In the $p_{\Lambda}$ region, more elaborate calculations, 
e.g. taking into account core-nucleus $1h\omega$ states, are needed to 
fully understand the new data. The Sacaly-Lyon model of the elementary 
process gives reasonable cross sections for the hypernucleus production 
which suggests that this model describes correctly the elementary 
process at forward kaon angles.\vspace{3mm}

\noindent {\small One of the authors (P.B.) wishes to thank the organisers 
for their kind invitation to this highly stimulating Conference. 
This work was supported by the Grant Agency of the Czech Republic, 
grant 202/08/0984.}

\end{document}